# Constraining global changes in temperature and precipitation from observable changes in surface radiative heating

Chirag Dhara

*Centre for Climate Change Research, Indian Institute of Tropical Meteorology, Pashan, Pune 411 008, India*

E: chirag.dhara@tropmet.res.in

## Abstract

Changes in the atmospheric composition alter the magnitude and partitioning between the downward propagating solar and atmospheric longwave radiative fluxes heating the Earth's surface. These changes are computed by radiative transfer codes in Global Climate Models, and measured with high precision at surface observation networks. Changes in radiative heating signify changes in the global surface temperature and hydrologic cycle. Here, we develop a conceptual framework using an Energy Balance Model to show that *first order* changes in the hydrologic cycle are mainly associated with changes in solar radiation, while that in surface temperature are mainly associated with changes in atmospheric longwave radiation. These insights are used to explain a range of phenomena including observed historical trends, biases in climate model output, and the inter-model spread in climate change projections. These results may help identify biases in future generations of climate models.

# Introduction

The energy balance at the Earth's surface plays a central role in shaping the planetary climate. Radiative heating of the surface by the absorption of solar and atmospheric longwave radiation at equilibrium is balanced by cooling through the emission of longwave radiation, and the turbulent fluxes of sensible and latent heat (see Figure 1). Longwave emission from the surface is related to the surface temperature by the Stefan-Boltzmann law, while the latent heat flux, which equals the precipitation heat flux at long time scales, drives the hydrological cycle. Thus, variations in the energy fluxes heating the surface must be associated with changes in the two major climate variables that are the surface temperature and precipitation.

At first glance, it may seem that climate change is associated with perturbations in only the *total* surface radiative heating, and that it may not matter whether the surface is heated by solar or terrestrial radiation. However, observational and modelling evidence suggest that this is not the case. For instance, observational networks were used to estimate that during 1960-1990 increasing atmospheric $CO_2$ concentrations increased downward longwave radiation at the surface by 3 W/m$^2$ but that anthropogenic aerosol loading in the atmosphere resulted in a solar dimming of nearly 10 W/m$^2$ (Wild *et al.*, 2004). Yet, in what might appear paradoxical, the study also estimated a rise in surface temperature by about 0.4°C in the same period despite the nearly 7 W/m$^2$ net reduction in surface heating.

Here, we study this problem from a surface energy perspective using a simple Energy Balance Model describing a climate in radiative-convective equilibrium (RCE). We develop a conceptual formalism to study the magnitude and partitioning between the energy fluxes cooling the surface, and quantify the changes in surface temperature and hydrological cycle associated with changes in the surface heating fluxes.

This formulation explains, *to first order*, a broad range of phenomena such as the inter-model spread among Global Climate Models (GCMs) in their historical and climate change simulations, observed trends in climate such as the "paradox" previously alluded to, and the propagation of GCM radiative biases (Wild *et al.*, 2013) into biases in temperature and precipitation (Mueller and Seneviratne, 2014). One of the important implications of our results is that large temperature and precipitation biases may persist even if there is no net change in total surface radiative heating. This study closes with general insights into bias propagation that apply across generations of climate models, including the next generation of GCMs (Phase 6 of the Coupled Model Inter-comparison Project (CMIP6)).

## Methods and data

The Earth's surface is heated by the net absorbed solar shortwave ($R_{sn}$) and downwelling longwave ($R_{l,d}$) radiation and cooled by the emission of longwave radiation ($R_{l,up}$) and the turbulent fluxes of sensible ($H$) and latent heat ($\lambda E$).

At equilibrium, the total heating and total cooling fluxes balance at the surface:

$$R_{l,up} + J = R_{sn} + R_{l,d}, \qquad (1)$$

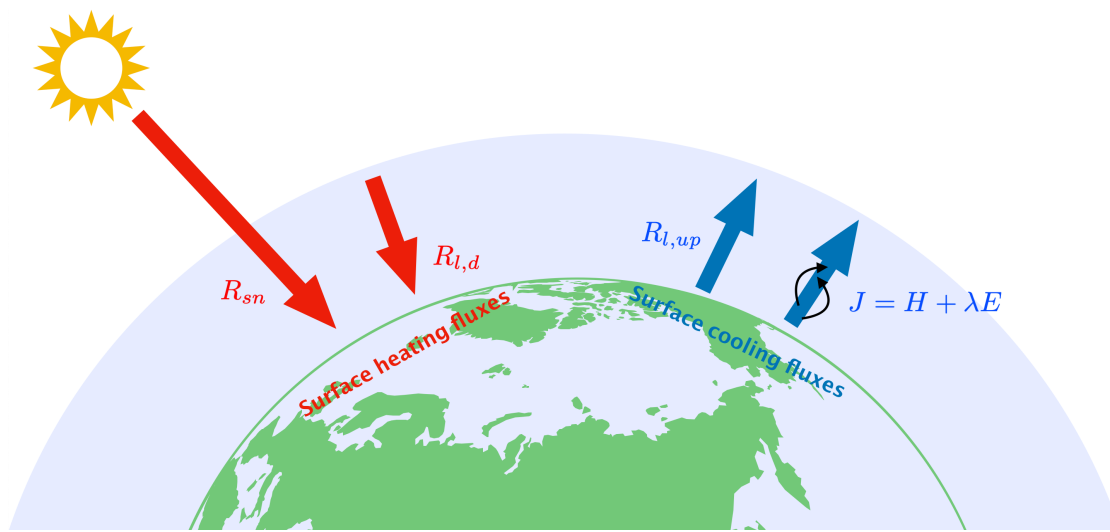

*Figure 1: Schematic of the Earth's equilibrium surface energy budget. The surface is heat by absorption of net solar radiation ($R_{sn}$) and longwave radiation ($R_{l,d}$) from the atmosphere. It is cooled by emission of longwave radiation ($R_{l,up}$), and the turbulent fluxes of sensible ($H$) and latent heat ($\lambda E$) from the surface.*

where $J = H + \lambda E$ is the convective flux. Heat storage and ground heat fluxes are assumed negligible at equilibrium. Of particular importance is the physical significance of the two cooling fluxes. Longwave emission from the surface is related to the surface temperature through the Stefan-Boltzmann law while the latent heat flux is the driver of the hydrological cycle since evaporation ($E$) equals precipitation ($P$) over long time scales.

Changes in the radiatively active constituents of the atmosphere such as greenhouse gases (GHGs) and aerosols modify the exchange of energy fluxes between the surface, atmosphere and space. The climate attains a new equilibrium over long time scales in which the surface heating fluxes are modified relative to the previous equilibrium state by $\Delta R_{sn}$ and $\Delta R_{l,d}$. Energy balance at the surface implies associated changes in the cooling fluxes $\Delta R_{l,up}$ and $\Delta J$ such that:

$$\Delta R_{l,up} + \Delta J = \Delta R_{sn} + \Delta R_{l,d}. \qquad (2)$$

Equation (2) is merely a restatement of the surface energy balance. However, our central result concerns the asymmetry in the surface fluxes $\Delta R_{l,up}$ and $\Delta J$ to the heating anomalies $\Delta R_{sn}$ and $\Delta R_{l,d}$ and the wide explanatory power of the relationships between the anomalies so-derived. Note that, net energy balance throughout the atmospheric column is implicit at equilibrium, although magnitudes of the constituent energy fluxes may vary across equilibrium states. The net solar energy absorption (equivalently, the outgoing longwave flux) at the Top of the Atmosphere (TOA) is assumed to be fixed.

We use a semi-analytic vertical column Energy Balance Model with a gray atmosphere (hereafter the "RC model") where the net upward and downward propagating energy fluxes balance throughout the atmosphere (Robinson and Catling, 2012). The upper atmosphere is assumed to be in radiative-only equilibrium while convection is assumed to ensue in the lower atmosphere when the radiative lapse rate exceeds the moist adiabatic lapse rate (Manabe and Strickler, 1964). The model is calibrated to present day conditions using data from the ERA-Interim reanalysis (Dee *et al.*, 2011). Also used in the analysis are simulation data from 22 Global Climate Models (GCMs) from Phase 5 of the Coupled Model Inter-comparison Project (Taylor, Stouffer and Meehl, 2012).

We use this model to study two hypothetical climate change scenarios by: 1. perturbing only the solar radiation absorbed at the surface (with no change in longwave heating) and 2. Perturbing only the longwave flux absorbed at the surface (no change in solar heating), and study the surface response in each case.

These scenarios are termed "hypothetical" since changes in the solar and longwave radiative fluxes are unlikely to occur independently in the real climate system (discussed further in the Discussion section). Yet, one of the major motivating factors for our use of an idealized model is that it allows us to treat the two scenarios as separable mathematical problems, and characterize their effects independently. Detailed methods are explained in the Supplementary Information (SI).

## Results

### The "surface-anomaly relationships"

We find that enhanced solar heating/cooling ($\sim \Delta R_{sn}$) (*e.g.* from aerosol perturbation) at the surface elicits mainly a convective ($\sim \Delta J$), thus hydrologic ($\sim \Delta \lambda E$), response. Conversely, enhanced longwave heating/cooling ($\sim \Delta R_{l,d}$) (*e.g.*

from GHG perturbation) elicits a response mainly in the longwave emission ($\sim \Delta R_{l,up}$), thus surface temperature ($\sim \Delta T_s$). In the context of the RC model, $\Delta \lambda E$ is deduced as the Priestley Taylor equilibrium evaporation flux (Priestley and Taylor, 1972) since it is not a standard output variable of the model (see SI).

The relationships between the anomalies *to first order* are described as the linear superposition of the two scenarios described in the Methods section, resulting in the following expressions (details in SI):

$$\begin{aligned} \Delta R_{l,up} &= 0.14\ \Delta R_{sn} + 0.74\ \Delta R_{l,d}, \\ \Delta J &= 0.86\ \Delta R_{sn} + 0.26\ \Delta R_{l,d}, \end{aligned} \quad (3)$$

from which the corresponding relationships for the two observables of interest are:

$$\begin{aligned} \Delta T_s &= 0.032\ \Delta R_{sn} + 0.138\ \Delta R_{l,d}, \\ \Delta \lambda E &= 0.72\ \ \Delta R_{sn} + 0.25\ \ \Delta R_{l,d}. \end{aligned} \quad (4)$$

The major insight, and the main thread underpinning all our results, is the asymmetry in these relationships, where unit changes in the solar and atmospheric longwave radiative fluxes (in $W/m^2$) affect temperature and precipitation differently.

We first focus on validation of Equations (4) since these constitute the most important climate variables. We use the historical, pre-industrial control (PI) and the abruptly quadrupled carbon dioxide (4xCO2) simulations from 22 CMIP5 GCMs. Magnitudes and errors in GCM output are computed as the long-term average of the global mean and standard deviation over the last 50 years to approximate equilibrium. For the historical dataset, we use the period 1956-2005.

Errors on the RC model are plotted in Figure 2 as a linear combination of the standard errors in GCM heating fluxes following Equations (4). In the SI, we demonstrate that the surface-anomaly relationships are robust to variations in the control climate used for calibration.

Anomalies for GCM variables are computed as follows:

1. for the historical simulation as the inter-model spread $\Delta X^i = X^i - \bar{X}$, where $X^i$ and $\bar{X}$ denote the global mean values for individual models and the multi-model mean, respectively.

2. for the climate change scenarios as $\Delta X^i = X^{4xCO2,i} - X^{PI,i}$ for each model.

$\Delta T_s$ and $\Delta P$ thus computed for the GCMs are plotted on the horizontal axes in Figure 2. The corresponding RC model derived values are plotted on the vertical axes, evaluating them from $\Delta R_s$ and $\Delta R_{l,d}$ for each GCM and using Equation (4), where $\Delta P^{RC}$ is computed as $\Delta \lambda E^{RC}/\lambda$.

## Spread among GCMs

In Figure 2a,b it is seen that the conceptual model explains about 65% of the inter-model spread in surface temperature among GCM historical simulations and 70% of the spread in precipitation. Despite the good correlation, it is clear from visual inspection that the explained variance and slope are affected by a single outlier model for each variable (*inmcm4* for surface temperature and *IPSL-CM5A-MR* for precipitation). Removing these outliers substantially improves the strength of the correlation to $r^2 = 0.84$ and $slope = 1$ for surface temperature and $r^2 = 0.72$ and $slope = 0.82$ for precipitation (figure not shown).

Figure 2c,d show the comparison of the RC model and GCMs for the change in climate between the pre-industrial control and abrupt 4xCO2 simulations. Our formalism explains nearly all the variance in GCM output for surface temperature

and nearly 75% of the variance in precipitation, although with a small systematic overestimation. Slopes of the linear regression are close to 1 for both variables.

GCM studies have found that precipitation response to GHG rise is the sum of an initial decrease due to a fast-response driven by the imposed radiative forcing and a subsequent increase via the temperature-mediated slow response (Bala, Caldeira and Nemani, 2010). However, the RC model only includes the temperature-mediated response, which may explain the systematic over-estimation seen in Figure 2d.

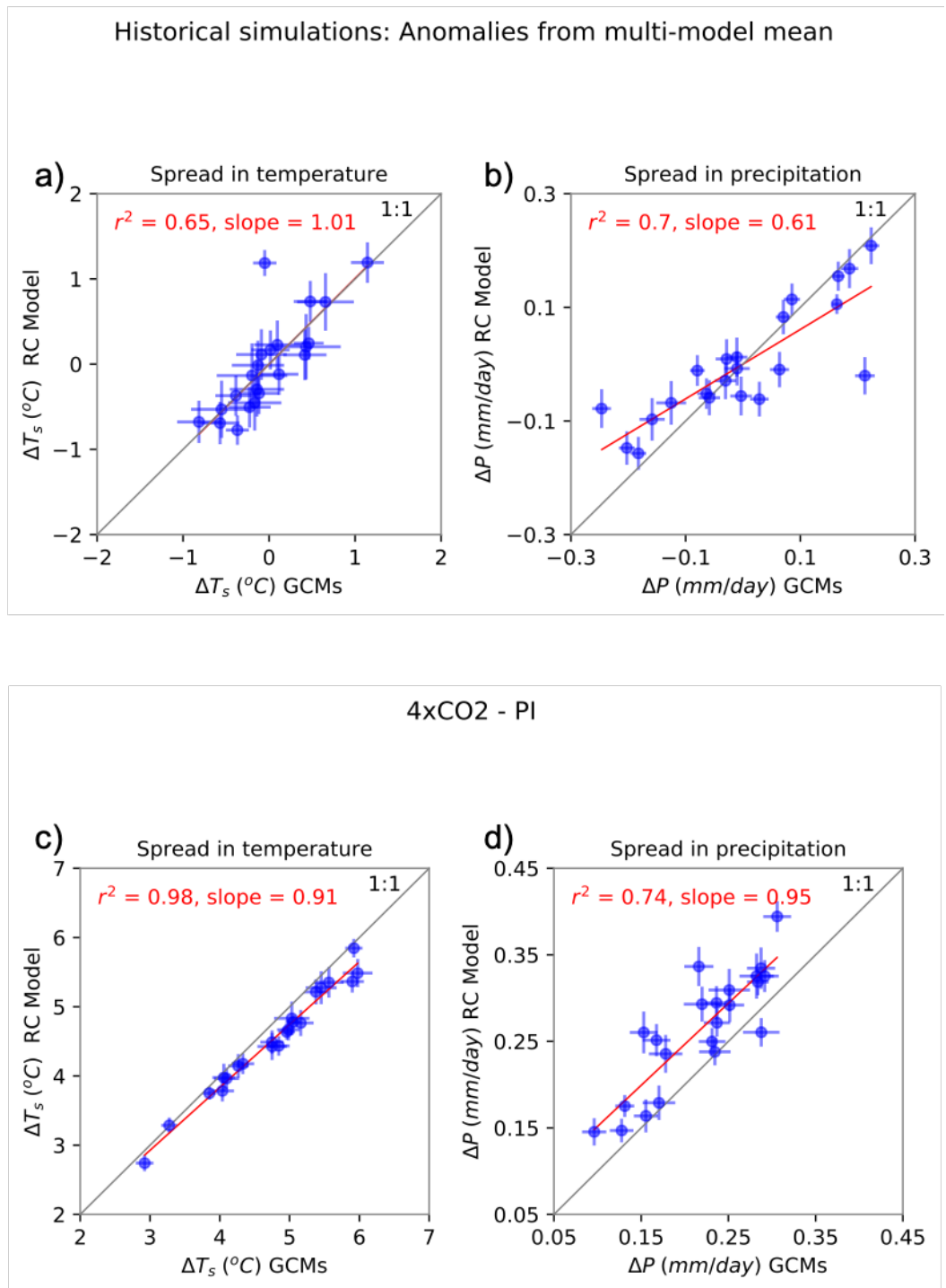

*Figure 2*: Analysis of spread among CMIP5 GCMs explained by the surface-anomaly relationships. Anomalies in (a) surface temperature and (b) precipitation w.r.t the multi-model mean in the GCM historical simulations. Change in the globally averaged (c) surface temperature and (d) precipitation (4xCO2 relative to the PI control) for the same GCMs.

These strong correlations suggest that the surface-anomaly relationships describe general constraints on the surface energy budget.

**Hydrological sensitivity**

Modification in the atmospheric greenhouse content manifests at the surface primarily as a change in the downward longwave flux (Ma, Wang and Wild, 2014). Assuming only a surface longwave perturbation ($\Delta R_{sn} = 0$), and using $\lambda E \approx 85$ W/$m^2$ as the estimate of the present day latent heat flux (Wild, 2017b) one can compute the hydrologic sensitivity as $\left(\frac{1}{\lambda E}\right) \frac{\Delta \lambda E}{\Delta T_s} \approx \left(\frac{1}{85}\right) \frac{0.25}{0.138} \approx 2.1 \ \% \ K^{-1}$ by simple substitution in Equation (4). This lies in the middle of the range of $1 - 3 \ \% \ K^{-1}$ simulated by GCMs (Held and Soden, 2006). The excellent agreement of our semi-empirical sensitivity estimate with detailed numeric simulations is likely because the global hydrological cycle is known to be limited by energy rather than moisture (Allen and Ingram, 2002).

**Explanation for GCM biases**

Equation (4) also expresses the propagation of biases in shortwave and longwave surface fluxes into biases in surface temperature and precipitation simulated by GCMs. Wild *et al.*, 2013 studied CMIP5 GCM simulations of the present-day surface heating fluxes using surface based observation stations. It was found that GCMs were systematically and significantly biased. The multi-model global mean bias in the downward longwave radiation was found to be $\Delta R_{l,d} = -6 \ W/m^2$ while in the absorbed shortwave radiation $\Delta R_{sn} = +10.5 \ W/m^2$. These heating biases are used in conjunction with Equation (4) to find the consequent bias in the climate variables (Table 1). We find that our estimates bear close resemblance to independent estimates of biases in GCM simulations of surface temperature and evapotranspiration studied in (Mueller and Seneviratne, 2014) using synthesized reanalysis datasets. Note that the latter estimates are over land only, and gridded datasets for GCM biases in evapotranspiration / precipitation with global coverage are not available to the author's knowledge.

**Table 1**: *GCM bias propagation from radiative heating fluxes to temperature and latent heat flux. In the top row are the GCM biases in the radiative fluxes, adopted from (Wild et al., 2013). These are used to compute biases in surface temperature and the latent heat flux from our surface-anomaly relationships (middle column). The data column (right) are estimates by Mueller and Seneviratne, 2014.*

| GCM biases (Wild *et al.*, 2013): $\Delta R_{sn} = +10.5 \ W/m^2$ and $\Delta R_{l,d} = -6 \ W/m^2$. | | |
|---|---|---|
| **Variables** | **Bias (Our estimates)** | **Bias (Mueller and Seneviratne, 2014)** |
| $\Delta T_s \ (^oC)$ | $-0.5$ | $-0.4$ |
| $\Delta \lambda E \ (W/m^2)$ | $+6.1$ | $+4.8$ |

**Observed trend in solar dimming and temperature rise**

Analyzing a system of surface observation stations, Wild *et al.*, 2004 estimated that from 1960-1990, the magnitude of shortwave radiation absorbed at the surface decreased by $-6 \ to -9 \ W/m^2$ ("solar dimming") whereas the downward longwave radiation increased by about +3 $W/m^2$. Despite the significant net reduction of surface heating $(\Delta R_{sn} + \Delta R_{l,d} = -3 \ to -6 \ W/m^2)$, surface temperature was found to have increased robustly by 0.4°C in the same period. As before, we use the observed changes in surface heating in Equation (4).

While there are discrepancies (**Table 2**), we find that the surface temperature indeed increases whereas it is the latent heat flux that is suppressed. The discrepancies seen may be partly related to the significant role of scattering aerosols in solar dimming (Ramanathan et al., 2001). Scattering results in a reduction in the net incoming solar radiation (i.e. a TOA anomaly), and not just surface-atmosphere energy redistribution that is assumed in formally deriving the surface-anomaly relationships. Yet, TOA anomalies ultimately modify the surface energy budget by perturbing the surface heating fluxes and the latter falls within the ambit of this formalism. Thus, one may expect that this formalism can partially explain the effect of aerosol scattering despite the TOA anomaly, with the net result being the quantitative discrepancy.

Despite the foregoing caveat, the surface-anomaly relationships not just capture the correct climate trends but offer an important physical insight that, to the author's knowledge, has not been previously stated as such: surface temperature could continue to increase because of its disproportionate sensitivity to the relatively small increase in surface longwave heating. In contrast, the latent heat flux being more sensitive to the reduction in solar heating was strongly suppressed.

*Table 2: Observational record of the trends in the surface radiative heating and cooling fluxes during 1960-1990 compared with estimates from Equation (4).*

| **Heating fluxes**: $\Delta R_{sn} = -6\ to\ -9\ W/m^2$ and $\Delta R_{l,d} = +3\ W/m^2$ (Wild *et al.*, 2004). | | |
|---|---|---|
| **Variables** | **(Wild *et al.*, 2004)** | **Our estimates** |
| $\Delta T_s\ (^oC)$ | $+0.4$ | $+0.13\ to\ +0.22$ |
| $\Delta \lambda E\ (W/m^2)$ | Unstated | $-3.6\ to\ -5.7$ |

### Rate of global warming in recent decades

Using worldwide observations of surface radiation (Driemel *et al.*, 2018), *Wild, 2017a* argued that downward longwave radiation has been increasing at a rate of about $+2 \, W/m^2$ per decade in recent decades due to increasing GHGs. Using these, we estimate a rate of change in temperature of about $+0.28^oC$ per decade, which is consistent with the latest estimate of $0.2 \pm 0.1^oC$ per decade as assessed in the recent IPCC SR1.5 (Allen *et al.*, 2018).

An important question that arises here is: why have the surface-anomaly relationships been employed above to explain transient changes although underpinned by an equilibrium model? Indeed, it is evident that these relationships cannot hold at very short – e.g. diurnal – time scales where ground heat flux and atmospheric heat storage contribute substantially to the energy budget.

While the foregoing analyses of observed trends may be considered transient, they may simultaneously be considered "quasi-equilibrium" changes in the sense of having analysed decadal changes/trends, where the assumptions underpinning our model are reasonably approximated. For instance, heat storage and ground fluxes are small at interannual and longer timescales, and precipitation balances evaporation. In a quasi-equilibrium transient climate, not just is $\Delta T_s$ in Equation (4) different from its final equilibrium value (say, $\Delta \mathrm{T}_{s,eq}$) but so are $\Delta R_{sn}$ and $\Delta R_{l,d}$ from their final equilibrium values ($\Delta R_{sn,eq}$ and $\Delta R_{l,d,eq}$). Therefore, we argue that the "quasi-equilibrium anomalies" also effectively co-vary as described by Equations (3) and (4).

### A generalized formulation for biases in observed climate variables

Since the surface-anomaly relationships of Equations *(3)* and *(4)* are derived from a conceptual model, their implications are expected to be independent of the detailed structure of complex coupled climate models.

The general propagation of biases in radiative heating to the climate variables of interest is shown in *Figure 3*. It is seen that biases in the heating fluxes ranging between $-6$ to $+6\ W/m^2$ can induce substantial biases in $T_s$ ranging from $-1$ to $+1\ ^oC$, and in $P$ from $-0.2$ to $+0.2\ mm/day$.

*Figure 3* also demonstrates that large biases can exist in the response variables even if there is no net change in surface heating *i.e.* $\Delta R_{sn} + \Delta R_{l,d} = 0$. For instance, $\Delta R_{sn} = -\Delta R_{l,d} = 4\ W/m^2$ results in the biases of $\Delta T_s \approx -0.4\ ^oC$ and $\Delta P \approx +0.07\ mm/day$.

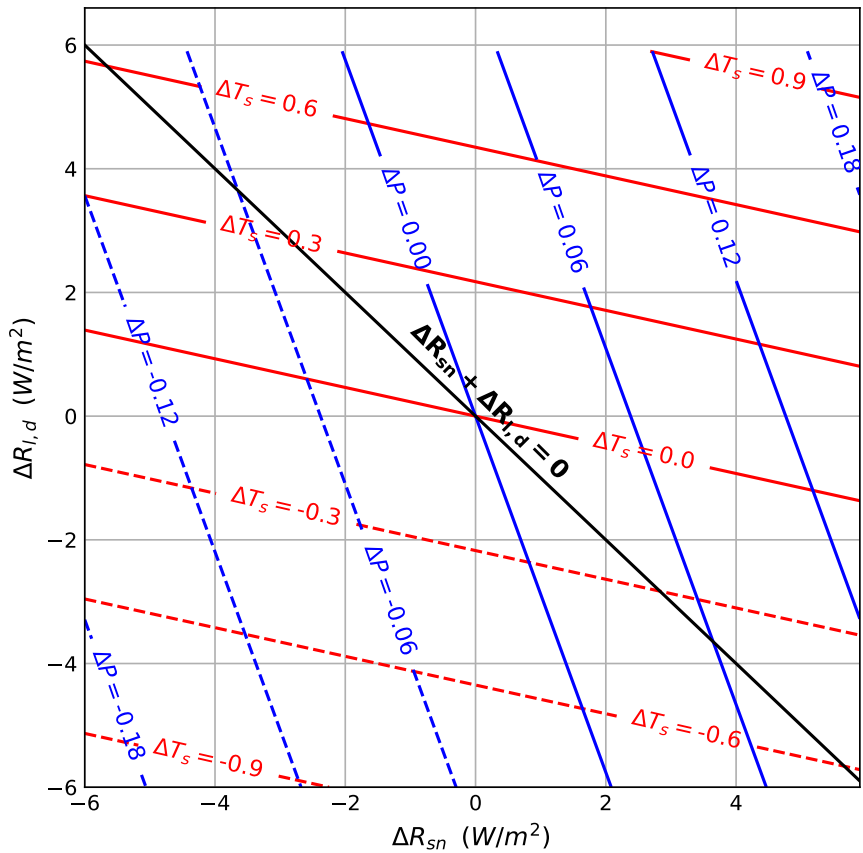

***Figure 3****: Propagation of anomalies (biases) in surface radiative heating fluxes into anomalies (bias) in surface temperature ($\Delta T_s$ ; $^oC$; red) and precipitation ($\Delta P$; $mm/day$; blue).*

We note that not all regions in this phase space may be sampled in the real atmosphere since the variations $\Delta R_{l,d}$ and $\Delta R_{sn}$ are likely to co-vary.

## Discussion

The highly idealized nature of underlying model naturally makes our results subject to several approximations and limitations. The approximations inherent to the RC model are discussed in Ramanathan and Coakley, 1978 and Robinson and Catling, 2012. Hence, we focus the discussion here to the two additional approximations made to derive the surface-anomaly relationships:

1. Our use of the two "hypothetical" climate change scenarios may suggest that this formulation carries an implicit assumption that the solar and longwave heating fluxes must vary independently. However, it is clear that these fluxes can co-vary in the atmosphere. For instance, an increase in atmospheric water vapour content, with an unperturbed TOA energy budget, would cause a decrease in the surface solar heating and a simultaneous increase in longwave heating.

However, here, our goal is to quantify the *first order* behavior of surface climate. In this context, we interpret use of the two hypothetical scenarios not as an assumption of independence but a mathematical linearization approximation, which is standard tool used in *first order studies* (Dhara, Renner and Kleidon, 2016).

2. The lower atmospheric lapse rate is held fixed in this formulation whereas it is known that the lapse rate may be modified with changes in climate (Hansen *et al.*, 2005). This assumption is made since the lapse rate is a specified parameter in the RC model and there are no additional physical constraints within this framework on how it may change with warming. However, previous *first order* studies have reported that the all-important water vapor feedback is well approximated under this assumption (Held and Soden, 2000). In addition, we demonstrate in the SI (Table 4) that the surface-anomaly relationships are robust to modifications in the assumed lapse rate.

One of the main limitations of this formalism derives from holding constant the equilibrium solar energy absorption at TOA to formally derive the surface-anomaly relationships. This makes their application most pertinent to analysing changes that affect mainly atmospheric *absorption*; these include changes in anthropogenic GHGs, water vapour and absorbing aerosols such as black carbon. On the other hand, changes in the concentration of scattering aerosols (Ramanathan *et al.*, 2001) and shortwave cloud feedbacks (Ceppi *et al.*, 2017) modify the TOA net solar energy absorption. Effects of these are only accounted for indirectly in this formalism by the perturbation of the surface heating fluxes by the TOA anomaly. Consequently, we anticipate that while the present formalism accounts partially for the effect of scattering on the surface variables, a more satisfactory description requires an expansion of this framework. This may be the subject of future work.

It is also important to recognize that it is neither the aim, nor is it possible to diagnose climate sensitivity from this work (Stocker *et al.*, 2013). While we have constrained changes in temperature and precipitation *given* the changes in surface heating fluxes, the latter are not (and cannot be) diagnosed apriory within this approach (Ramaswamy *et al.*, 2019).

Despite these limitations, the major merit of this approach is that it constitutes a process-agnostic conceptual framework to study changes in surface climate, that may otherwise get obscured by complex details such as the spectral properties of atmospheric molecules (Rothman *et al.*, 2009). Furthermore, changes in both the surface radiative heating fluxes are directly measurable through surface observation networks such as the Global Energy Balance Archive (GEBA) (Gilgen and Ohmura, 1999; Wild *et al.*, 2017) and the Baseline Surface Radiation Network (BSRN) (Driemel *et al.*, 2018). Thus, our formulation allows a direct inference of the variations in temperature and precipitation, *to first order*, for measured changes in surface radiation. This is particularly noteworthy given the paucity of globally representative observations of turbulent fluxes at the surface (Wild, 2017b).

A plausible physical mechanism for the seemingly counter-intuitive asymmetry in the surface response variables to short- and long wavelength heating fluxes is the differing potential of these fluxes to generate atmospheric instability. Solar absorption at TOA being held fixed in our formulation, an increase in surface shortwave absorption comes at the expense of atmospheric absorption, resulting in greater atmospheric instability and prompting a stronger hydrologic response. Conversely, increased longwave heating of the surface occurs concurrently with an increasing absorption of longwave radiation (e.g. because of increasing GHGs) in the atmosphere. Thus, the dominant surface response is in surface temperature rather than precipitation.

An extension of these results may lend important insight into other important climate forcings, particularly those that directly affect surface properties, such as land use land cover changes (Davin, de Noblet-Ducoudré and Friedlingstein, 2007) or the intensification of irrigation practices (Boucher, Myhre and Myhre, 2004).

## Acknowledgements

I thank the Coupled Model Inter-comparison Project for the GCM simulation data used in this work. These data are available at https://esgf-data.dkrz.de/search/esgf-dkrz/. I also thank the European Centre for Medium-Range Weather Forecasts for the reanalysis data used here These may be downloaded from: https://www.ecmwf.int/en/forecasts/datasets/reanalysis-datasets/era-interim. I acknowledge that the work was initiated at the Max Planck Institute for Biogeochemistry, Jena, Germany and thank Axel Kleidon and Maik Renner for suggestions and comments on the manuscript.